\documentclass[aps,prl,superscriptaddress,showpacs,showkeys,floatfix,reprint]{revtex4-1}
\usepackage{graphicx}
\usepackage{dcolumn}
\usepackage[pdftex]{color}
\usepackage{amsmath}
\usepackage{caption}
\usepackage{subcaption}
\usepackage{bm}
\usepackage{gensymb}
\usepackage{verbatim}
\definecolor{darkgreen}{rgb}{0.133,0.545,0.133}
\definecolor{orange}{rgb}{1.0,0.76,0.02}

\begin{document}

\title{First principles characterization of reversible martensitic transformations}

\author{Alberto Ferrari}
\email{alberto.ferrari@rub.de}
\affiliation{Interdisciplinary Centre for Advanced Materials Simulation, Ruhr-Universit{\"a}t Bochum, 44801 Bochum, Germany}
\author{Davide G.~Sangiovanni}
\affiliation{Interdisciplinary Centre for Advanced Materials Simulation, Ruhr-Universit{\"a}t Bochum, 44801 Bochum, Germany}
\affiliation{Department of Physics, Chemistry, and Biology (IFM), Link{\"o}ping University, SE-58183 Link{\"o}ping, Sweden}
\author{Jutta Rogal}
\affiliation{Interdisciplinary Centre for Advanced Materials Simulation, Ruhr-Universit{\"a}t Bochum, 44801 Bochum, Germany}
\author{Ralf Drautz}
\affiliation{Interdisciplinary Centre for Advanced Materials Simulation, Ruhr-Universit{\"a}t Bochum, 44801 Bochum, Germany}

\date{\today}

\begin{abstract}
Reversible martensitic transformations (MTs) are the origin of many fascinating phenomena, including the famous shape memory effect.
In this work, we present a fully \textit{ab initio} procedure to characterize MTs in alloys and to assess their reversibility.
Specifically, we employ \textit{ab initio} molecular dynamics data to parametrize a Landau expansion for the free energy of the MT.
This analytical expansion makes it possible to determine the stability of the high- and low-temperature phases, to obtain the Ehrenfest order of the MT, and to quantify its free energy barrier and latent heat.
We apply our model to the high-temperature shape memory alloy Ti-Ta, for which we observe remarkably small values for the metastability region (the interval of temperatures in which the high- and low-temperature phases are metastable) and for the barrier:
these small values are necessary conditions for the reversibility of MTs and distinguish shape memory alloys from other materials.
\end{abstract}



\maketitle

A martensitic transformation (MT)~\cite{Ashby1998} is a diffusionless phase transition, triggered by temperature or stress, that changes the symmetry of a high-temperature phase (austenite) and forms variants of a low temperature phase (martensite).
Most of the MTs are irreversible, as dislocations, shear, and plastic deformation accumulate during the transformation.
However, if the symmetry of martensite is lower than that of austenite and if the variations in lattice parameters and atomic volumes are small, the MT can be reverted, that is, the system can be switched between the two phases with small latent heat~\cite{Bhattacharya2004,Bhattacharya2005,James2005,Cui2006}.
Reversible MTs in metals or polymers are appealing as they often result in the shape memory effect, the ability to recover a predetermined shape upon heating, and pseudoelasticity, the capacity to accommodate large deformations without plasticity~\cite{Chang1951,Lendlein2001,Jani2014}.
Other examples in which reversible MTs are important include the recently discovered gum metals~\cite{Hao2013}, where metastable phases have been observed to form via reversible transformations~\cite{Zhang2017}.

An urgent technological challenge for actuator and biomedical applications is to identify alloys that exhibit reversible MTs that are stable during operational cycles.
With very few exceptions~\cite{Haskins2016}, first principles investigations aiming to clarify the mechanisms underlying a MT generally rely on static, $T=0$~K calculations.
These, however, are often inadequate to describe the atomistic processes responsible for the dynamic and/or thermodynamic stabilization of the austenite phase at finite temperatures, as well as the interval of temperatures in which austenite and martensite are metastable (metastability region), the free energy barrier, the latent heat, and even the Ehrenfest order of a MT.  

To overcome these limitations we have employed \textit{ab initio} molecular dynamics (aiMD) simulations to access structural properties at finite temperature, and combined our \textit{ab initio} data with a 2-4-6 Landau-Falk expansion of the free energy~\cite{Falk1980,Khalil-Allafi2005} to characterize the nature of reversible MTs and suggest necessary conditions to distinguish them from irreversible ones.
We have applied our method to the shape memory alloy Ti-Ta~\cite{Bagaryatskii1958,Bywater1972,Fedotov1985,Buenconsejo2009,Niendorf2015,Chakraborty2015,Chakraborty2016,Kadletz2017,Kadletz2018,Ferrari2018} that features a reversible MT with a high ($>$100\degree C) transition temperature.
Our key findings include that, in this system, there is only a small interval of temperatures where austenite and martensite are both dynamically stable and that, in this interval,
the two phases are separated by an extremely small free energy barrier.

Any first order phase transition, like the reversible MT described here, involves the nucleation and growth of a new phase inside the other;
the consideration of this mechanism is beyond the scope of this work.
Nevertheless, even for a homogeneous transition, small metastability regions, energy barriers and latent heats generally distinguish reversible MTs from ordinary MTs;
with our approach we provide a fully \textit{ab initio} strategy to identify these fundamental characteristics of a MT.

\begin{figure}
\begin{centering} 
\includegraphics[scale=0.15]{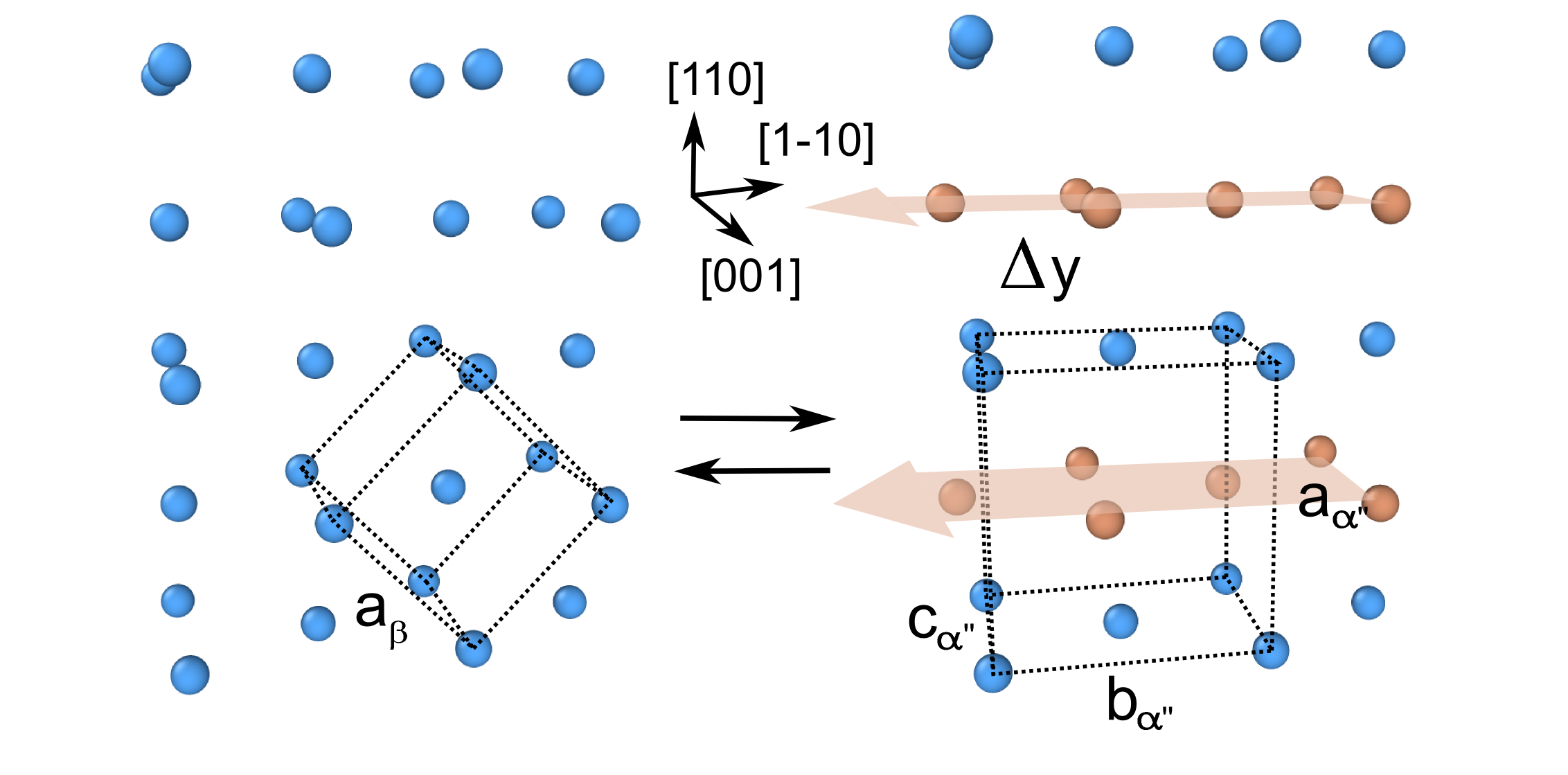}
\par\end{centering}
\caption{The MT in Ti-Ta. 
The austenitic phase (\textbf{left}) is a bcc structure. 
The martensitic phase (\textbf{right}) is orthorhombic, and it is obtained from the austenitic phase by cell distortion and gliding of alternating $\left\{ 110\right\}$ planes (in brown) along the $\left\langle -110\right\rangle$ direction, described by the parameter $\Delta y$.
}
\label{bet_alp}
\end{figure}
\begin{figure}
\begin{centering} 
\includegraphics[scale=0.20]{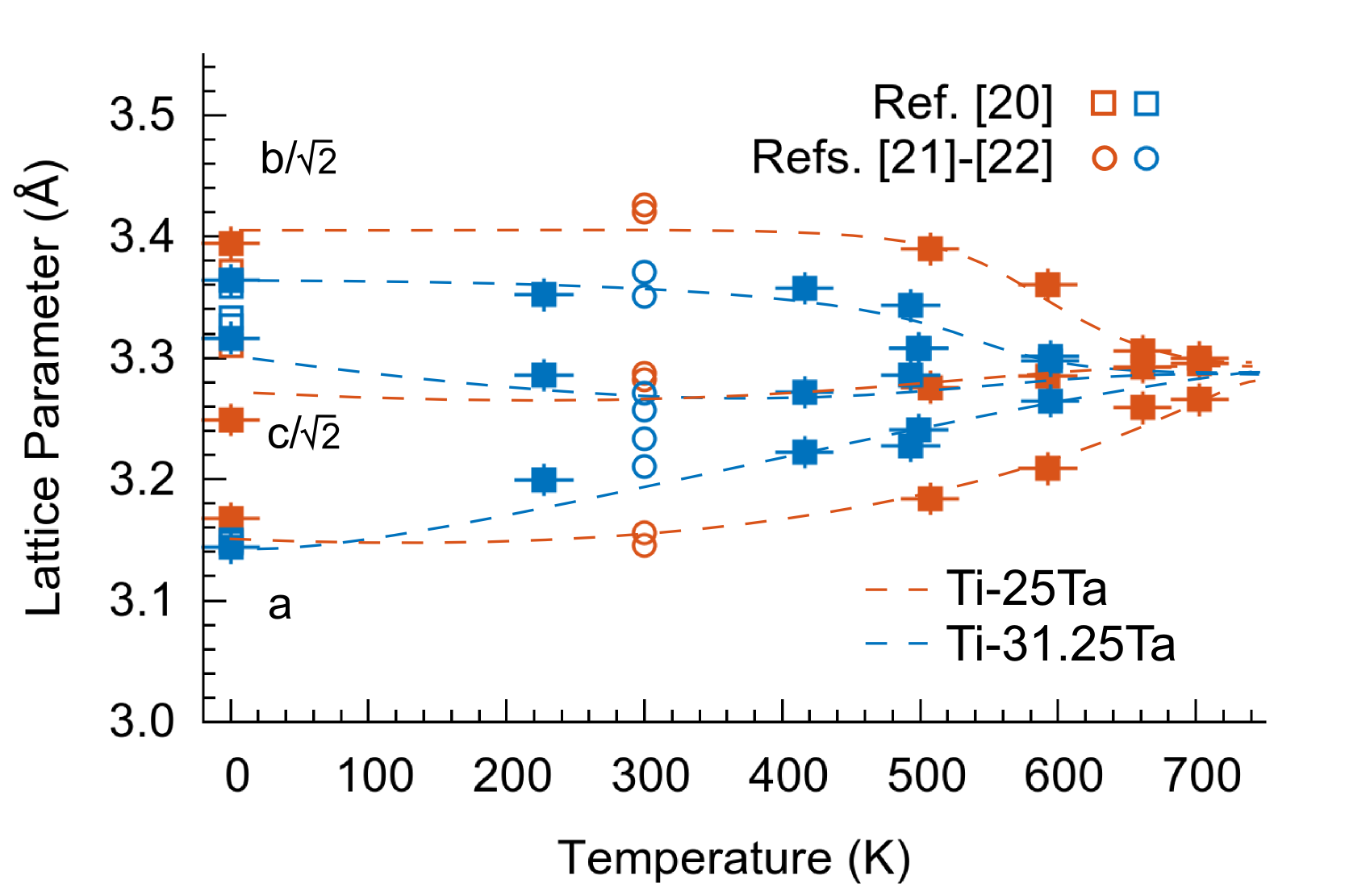}
\par\end{centering}
\caption{Lattice parameters of Ti-25Ta (red) and Ti-31.25Ta (blue) as a function of temperature. 
Circles are experimental data on bulk samples and thin films at room temperature. 
Empty squares are DFT calculations from Ref.~\cite{Chakraborty2016}. 
Broken lines are guide-to-the-eye.}
\label{lat}
\end{figure}
\begin{figure*}[t]
\begin{centering} 
\includegraphics[scale=0.20]{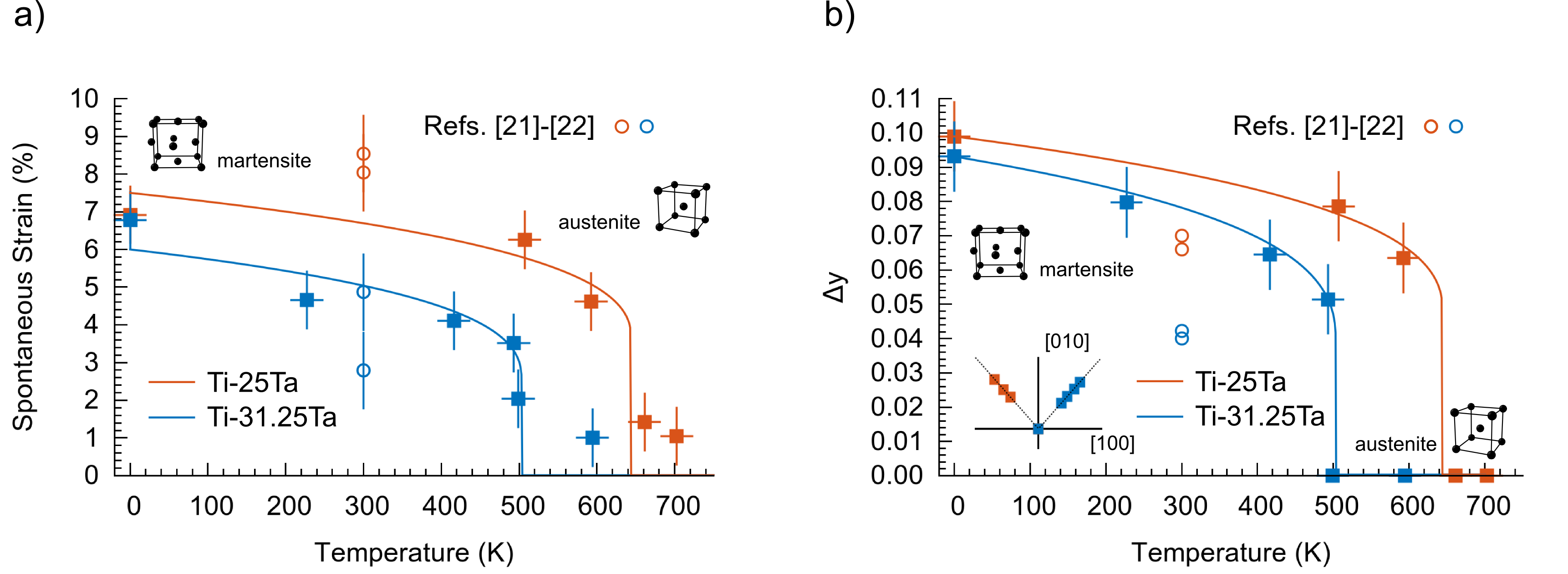}
\par\end{centering}
\caption{
\label{sls_and_y}
a) Spontaneous lattice strain of $\alpha''$ for Ti-25Ta (red) and Ti-31.25Ta (blue) as a function of temperature. 
b) The time- and atom-averaged  $\Delta y$ parameter as a function of temperature for Ti-25Ta (red) and Ti-31.25Ta (blue). The inset shows the directions of the average atomic displacements observed in the aiMD simulations.
For both order parameters, squares are extracted from the aiMD simulations, circles are experimental data for bulk and thin film samples at room temperature, and solid lines are predictions from the Landau-Falk expansion (no fit).  
}
\end{figure*}
\begin{figure}[b]
\begin{centering} 
\includegraphics[scale=0.20]{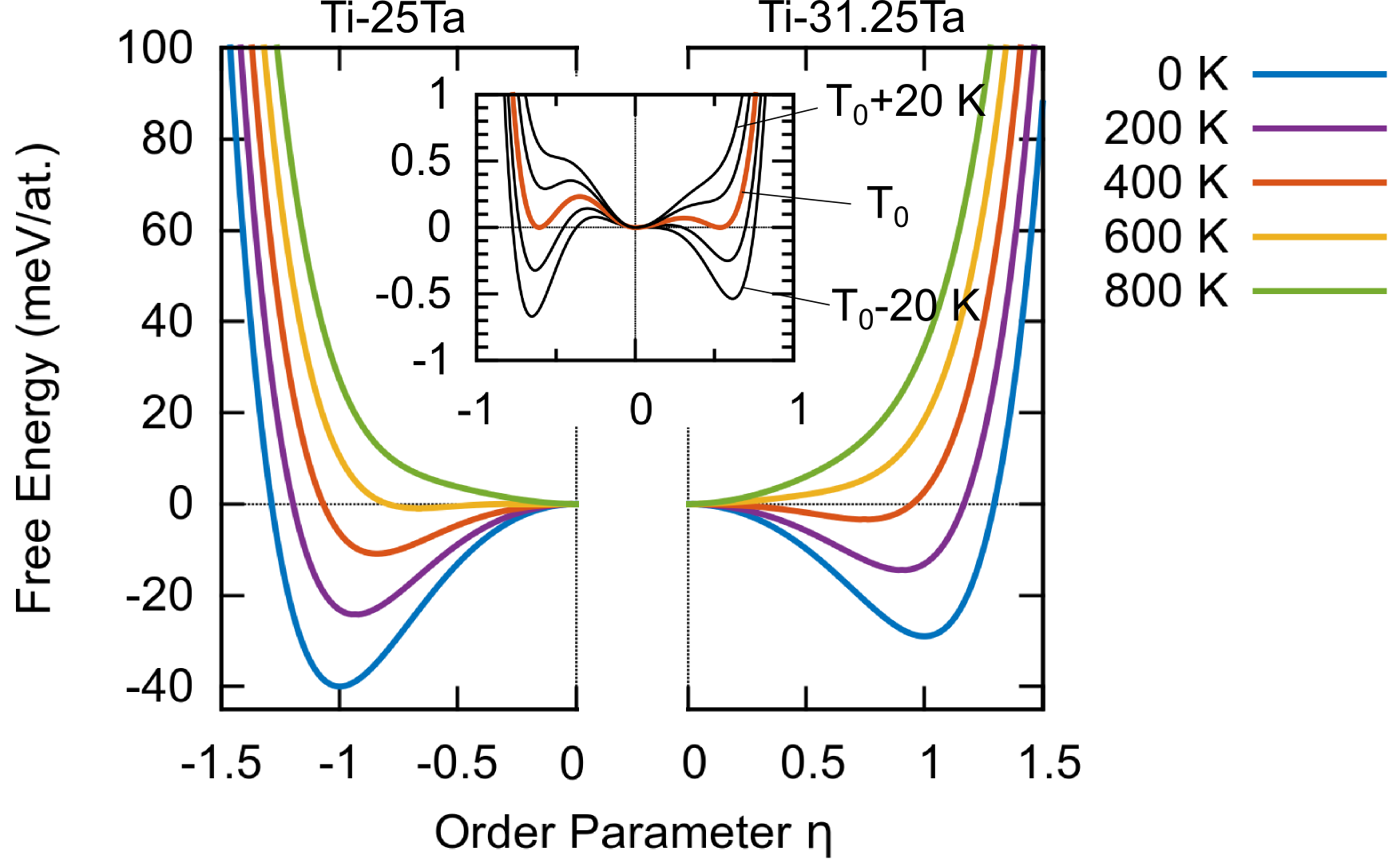}
\par\end{centering}
\caption{The free energy profiles as a function of the order parameter for Ti-25Ta (\textbf{left}) and Ti-31.25Ta (\textbf{right}) for different temperatures. The inset shows the same profiles for an interval of temperatures around the transition temperature $T_0$.}
\label{free_en}
\end{figure}
%

The austenitic phase in Ti-Ta is a solid solution of Ti and Ta with body-centered cubic (bcc) symmetry, called $\beta$ phase.
At lower temperatures the $\beta$ phase breaks its cubic symmetry and transforms into one of the twelve-fold degenerate orthorhombic martensitic variants, called $\alpha''$.
As depicted in Fig.~\ref{bet_alp}, $\alpha''$ (right panel) is obtained from $\beta$ (left panel) by an orthorhombic cell distortion and a displacement of alternating $\left\{ 110\right\}$ atomic planes along $\left\langle -110\right\rangle$ directions.
The lattice vectors of the martensitic phase are ${(a_{\alpha''},0,0), (0,b_{\alpha''},0), (0,0,c_{\alpha''})}$, with $a_{\alpha''}<c_{\alpha''}/\sqrt{2}<b_{\alpha''}/\sqrt{2}$.
The MT in Ti-Ta can be described by two order parameters that change together: the spontaneous lattice strain (SLS) of martensite, which accounts for the respective elongation and shrinkage of the lattice parameters, and the average displacement from ideal bcc positions $\Delta y$.
The SLS is given by~\cite{Kadletz2018}:
\begin{equation}
\text{SLS}= 2 \cdot \frac{b_{\alpha''}/\sqrt{2}-a_{\alpha''}}{b_{\alpha''}/\sqrt{2}+a_{\alpha''}} \quad,
\label{sls}
\end{equation}
%
and $\Delta y$ is the average relative distance of the atoms in the gliding planes from the ideal bcc positions.

We have performed Parrinello-Rahman~\cite{Parrinello1980} aiMD simulations in the $NPT$ ensemble using special quasirandom structures (SQS)~\cite{Zunger1990} for two compositions with 25~at.\% and 31.25~at.\%~Ta (Ti-25Ta and Ti-31.25Ta, see the Supplemental Material \cite{SuppMat} for the details of the calculations).
SQSs arrangements mimic solid solutions by minimizing geometrical $n$-body correlations.
For Ti-25Ta we have carried out aiMD simulations at 500~K, 600~K, 650~K, and~700 K, whereas for Ti-31.25Ta at 230~K, 415~K, 500~K, and 600~K.

In Fig.~\ref{lat} the average lattice parameters $a$, $b/\sqrt{2}$, and $c/\sqrt{2}$ extracted from the aiMD simulations are presented as a function of temperature, and compared to previous $T=0$~K calculations~\cite{Chakraborty2016} and experimental data on bulk samples~\cite{Kadletz2017} and thin films~\cite{Kadletz2018}.
At low temperature the structures correspond to the orthorhombic $\alpha''$ phase, as $a<c/\sqrt{2}<b/\sqrt{2}$ for both compositions.
Our 0~K relaxed lattice constants are generally in very good agreement with the values by Chakraborty~\textit{et al.}~\cite{Chakraborty2016}, and the aiMD simulation results compare very well with the experimental data at room temperature by Kadletz~\textit{et al.}~\cite{Kadletz2017, Kadletz2018}.
At $T>600$ K and $T\geq500$ K for Ti-25Ta and Ti-31.25Ta, respectively, $b$ and $c$ become equal, indicating that the austenitic phase forms.
The fact that the lattice parameter $a$ is slightly smaller than $b/\sqrt{2}$ and $c/\sqrt{2}$ even at high temperatures, when the system is in the austenitic phase, is due to finite size effects.

The results for the SLS from the numerical simulations are shown as square symbols in Fig.~\ref{sls_and_y}a. 
The values of the calculated SLS are consistent with the experimental data.
At high temperatures the residual SLS is around 1\%, suggesting that the mentioned size effects are small. 

The square symbols in Fig.~\ref{sls_and_y}b represent the atomic displacements $\Delta y$, averaged over time and over all atoms in the supercell, as a function of temperature.
For both compositions, at low temperature the value of $\Delta y$ is approximately 0.1.
At $T>600$ K and $T\geq500$ K for Ti-25Ta and Ti-31.25Ta, respectively, $\Delta y$ drops to zero, which indicates that the average atomic positions coincide with those of an ideal bcc lattice.
The inset of Fig.~\ref{sls_and_y}b shows that the displacements for both compositions are in the $\left\langle1 1 0\right\rangle$ direction, consistent with the mechanism depicted in Fig.~\ref{bet_alp}.
The deviation of the theoretical $\Delta y$ values from the experimental data is attributed to the presence of phase separation in both the bulk and thin film samples in the experiments~\cite{Ferrari2018a}. 
Phase separation implies that the Ta content in the $\alpha''$ phase is considerably higher than the nominal composition of the samples and leads to a severe underestimation of the  $\Delta y$ value. 

From the temperature dependence of the two order parameters in our aiMD simulations the transition temperatures $T_0$ for Ti-25Ta and Ti-31.25Ta have been determined~\cite{SuppMat} to be approximately 625~K and 500~K, respectively, slightly overestimated in comparison to the experimental data (560~K and 420~K, respectively)~\cite{Ferrari2018}.
An even more severe overestimation has been noted before in aiMD simulations of the shape memory alloy NiTi~\cite{Haskins2016} and imputed to the absence of crystal defects and internal stresses in the calculations.
Our values should therefore be considered as an upper limit for $T_0$ in an ideal, defect free crystal.
As an additional possible source of error, the finite size of the simulation cell may induce artificial correlations.


To fully characterize the MT $\alpha''\rightleftharpoons\beta$ we can parametrize the free energy $F(V,T)$, which, at zero pressure, governs the thermodynamics of the phase transition.
For reversible MTs, Falk~\cite{Falk1980} has suggested a 2-4-6 Landau expansion of $F(V,T)$ as a function of a one dimensional order parameter $\eta$
\begin{equation}
 F(\eta,T)=a \eta^6 - b \eta^4+ c (T-T_c) \eta^2,
 \label{falk_free_en}
\end{equation}
%
where $a$, $b$, and $c$ are material-dependent parameters, and $T_c<T_0$ is the temperature at which the austenitic phase becomes metastable.
In this picture, $T_0$ is the temperature at which the free energies of austenite and martensite are equal.
\newline
In the case of the MT in Ti-Ta, Eq.~\eqref{falk_free_en} provides a one-dimensional description of the relative stability of austenite and one of the twelve-fold degenerate martensitic variants;
$\eta$ can be either the SLS or $\Delta y$, as in the MT the lattice constants and atomic positions are observed to change together.


Traditionally, Eq.~\eqref{falk_free_en} has been used to fit order parameters and latent heats measured experimentally. 
Here, we determine the parameters $a$, $b$, $c$, and $T_c$ exclusively from first principles simulation data.
Specifically, we have parametrized the free energy to reproduce the energy difference between $\beta$ and $\alpha''$ at 0~K, the transition temperature $T_0$, and the values of the order parameters at 0~K and at $T_0$ (see the Supplemental Material~\cite{SuppMat} for details).
The obtained free energy curves as a function of $\eta$ are presented in Fig.~\ref{free_en} for Ti-25Ta and Ti-31.25Ta at different temperatures.
At 0~K the austenitic phase (corresponding to $\eta=0$) is a maximum of the energy, whereas the martensitic phase (corresponding to $\eta=\pm 1$) is a minimum.
At this temperature there is no barrier separating the two states, meaning that austenite is unstable, in agreement with previous 0~K static calculations~\cite{Chakraborty2015}.
As the temperature increases, the martensitic minimum shifts towards smaller values of $\eta$.
At high temperature the free energy has only one minimum at the austenitic phase, hence the martensite is unstable.
The martensitic and austenitic phases are therefore found to be unstable in a very wide range of temperatures.
This is confirmed by our aiMD simulations: as initial configurations we used both the $\alpha''$ as well as the $\beta$ phase and apart from the simulations for Ti-31.25Ta at $T=500$~K the structure immediately transformed to the thermodynamically stable one, reflecting the instability of the corresponding other phase.

Within  the Landau-Falk expansion, however, a small interval of temperatures around $T_0$ is predicted in which both phases are metastable,  separated by a very small free energy barrier, as shown in the inset of Fig.~\ref{free_en}.
Consequently, the phase transition $\alpha''\rightleftharpoons\beta$ is of first order, in agreement with experiments~\cite{Kadletz2018}.
This is also supported by the numerical data: for Ti-31.25Ta at $T\sim$ 500 K we have found that the martensitic and austenitic phases coexist.
The presence of this free energy barrier is due to entropy contributions to the free energy and cannot be detected with 0 K calculations.
Finite temperature simulations are thus essential to capture the correct mechanism of stabilization of the austenitic phase.
In particular, the entropy difference $\Delta S$ between austenite and martensite induces a finite latent heat $T_0 \Delta S$ of the MT. 
We obtain from the Falk-Landau model values of $T_0 \Delta S=$ 19$\pm$3 meV/at.~ and 11$\pm$3 meV/at. for Ti-25Ta and Ti-31.25Ta, respectively.

Most notably, we extract from the analytical expansion metastability regions of only 70$\pm$30~K and 30$\pm$10~K, and free energy barriers of only 200$\pm$70 $\mu$eV/at.~and 100$\pm$30 $\mu$eV/at.~for Ti-25Ta and Ti-31.25Ta, respectively.
These exceptionally small values indicate that the MT in Ti-Ta is highly reversible.
In fact, such small metastability regions and energy barriers for bulk material are necessary properties that distinguish reversible MTs from irreversible MTs. 
For comparison, the energy barriers for the MTs in Fe-C alloys range between $20-50$~meV/at.~\cite{Zhang2015,Zhang2016}, which is approximately 2~orders of magnitude larger than the barriers we observe in Ti-Ta. 

A very small  free energy barrier is also consistent with our numerical calculations, as for Ti-31.25Ta we have captured a MT within one aiMD run (see Fig.~5 in the Supplemental Material~\cite{SuppMat}).
Another  factor that favors the reversibility of the MT is a small  difference in atomic volume between the martensite and austenite~\cite{Cui2006}, which is also fulfilled in Ti-Ta (details are given in the Supplemental Material~\cite{SuppMat}).  

The analytical expansion in Eq.~\eqref{falk_free_en} can further be used to extract the temperature dependence of the order parameters SLS and $\Delta y$: 
the value of the order parameter at each temperature is the one that minimizes the free energy at that particular temperature~\cite{SuppMat}.
The corresponding trends in SLS and $\Delta y$ predicted by the Landau-Falk expansion are presented in Fig.~\ref{sls_and_y} as solid lines.
The agreement between the aiMD data and the analytical predictions is remarkable.  We would like to stress that the parameters entering Eq.~\eqref{falk_free_en} have not been obtained by fitting the temperature dependence of the order parameters SLS and $\Delta y$, but  have been extracted from our first principles data at 0~K and $T_0$.
Furthermore, within the Landau-Falk expansion the two order parameters are predicted to be discontinuous at the transition temperature, confirming the first-order character of the MT.

In conclusion, we have successfully applied a combination of \textit{ab initio} molecular dynamics simulations with an analytical expansion of the free energy to characterize the most significant properties of martensitic transformations, which often cannot be captured by 0~K calculations. 
The methodology presented in this work is based entirely on first principles data and is very well suited to study  MTs in a variety of compounds.
In particular, we have applied this formalism to the technologically relevant Ti-Ta alloy, for which we have predicted for bulk transformations very small metastability regions (tens of K) and very small free energy barriers (hundreds of $\mu$eV).
These two quantities are decisive in specifying reversible MTs and have to be considered as the fundamental origin of the shape memory effect.

The work presented in this letter has been financially supported by the Deutsche Forschungsgemeinschaft (DFG) within the research unit FOR 1766 (High Temperature Shape Memory Alloys, http://www.for1766.de), under the grant number RO3073/4-2 (sub-project 3).
D.G.S.~acknowledges financial support from the Olle Engkvist Foundation.
The computations have been performed using the Gamma and Triolith clusters, managed by the Swedish National Infrastructure for Computing (SNIC) at the National Supercomputer Centre (NSC) in Link{\"o}ping, the Kebnekaise cluster at the High Performance Computing Center North (HPC2N) in Ume\r{a}, and the Beskow cluster at the Center for High Performance Computing (PDC) in Stockholm.

\bibliography{./bib/biblio.bib}

\end{document}


\title{First principles characterization of reversible martensitic transformations \\ (Supplemental Material)}

\author{Alberto Ferrari}
\email{alberto.ferrari@rub.de}
\affiliation{Interdisciplinary Centre for Advanced Materials Simulation, Ruhr-Universit{\"a}t Bochum, 44801 Bochum, Germany}
\author{Davide Sangiovanni}
\affiliation{Interdisciplinary Centre for Advanced Materials Simulation, Ruhr-Universit{\"a}t Bochum, 44801 Bochum, Germany}
\affiliation{Department of Physics, Chemistry, and Biology (IFM), Link{\"o}ping University, SE-58183 Link{\"o}ping, Sweden}
\author{Jutta Rogal}
\affiliation{Interdisciplinary Centre for Advanced Materials Simulation, Ruhr-Universit{\"a}t Bochum, 44801 Bochum, Germany}
\author{Ralf Drautz}
\affiliation{Interdisciplinary Centre for Advanced Materials Simulation, Ruhr-Universit{\"a}t Bochum, 44801 Bochum, Germany}


\maketitle




\section{Computational Details}

\begin{figure}
\begin{centering} 
\includegraphics[scale=0.17]{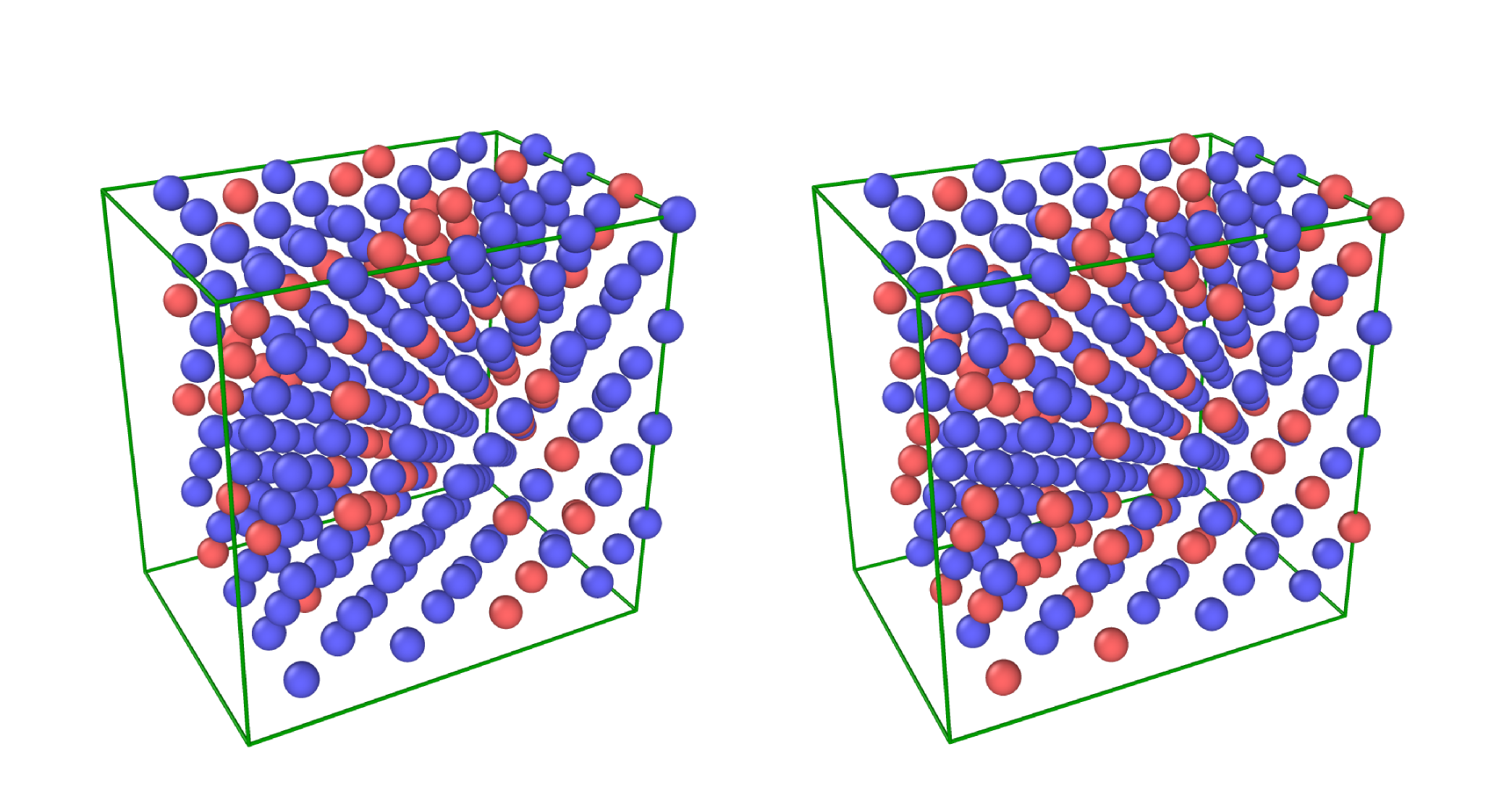}
\par\end{centering}
\caption{The SQS employed for the simulations of Ti-25Ta (\textbf{left}) and Ti-31.25Ta (\textbf{right}). Ti atoms are blue, and Ta atoms are red.}
\label{SQS}
\end{figure}

We have performed Parrinello-Rahman $NPT$ molecular dynamics~\cite{Parrinello1980,Parrinello1981} with a Langevin thermostat and an Andersen barostat with Langevin friction~\cite{Allen1991}, as implemented in the Vienna \textit{Ab initio} Simulation Package (VASP 5.4)~\cite{Kresse1993,Kresse1996,Kresse1996a}.
The friction coefficients of both the Langevin thermostat and the barostat have been set to $\gamma=0.1$ ps$^{-1}$, while a value of $M=1$ a.m.u.~has been used for the mass of the extended particle in the Andersen barostat.
With these settings, the root mean squared deviation of the instantaneous $T$ and $P$ from their average values was of the order of 20~K and 100~MPa, respectively.
A timestep of 1 fs has been employed for all simulations.
The sampling has always been started after complete equilibration of both temperature and pressure, and thermodynamic averages have been performed on trajectories with a duration of at least 7 ps.
\newline
Total energies and forces have been computed using density functional theory (DFT) with projector-augmented wave (PAW)~\cite{Bloechl1994,Kresse1999} pseudopotentials including $s$, $p$, and $d$ electrons for Ti and Ta.
The generalized gradient approximation (GGA) functional parametrized by Perdew, Burke, and Ernzerhof (PBE)~\cite{Perdew1996} has been utilized for the exchange-correlation term.
To integrate the Brillouin zone, we have employed the Monkhorst-Pack scheme~\cite{Baldereschi1973,Monkhorst1976} with a k-point mesh with a linear density of 0.3 2$\pi/$\AA.
The electronic occupations have been smeared with the Methfessel-Paxton method~\cite{Methfessel1989} with a width of 0.05~eV.
The energy cutoff has been fixed to 400~eV.
These settings have been found to ensure an accuracy of approximately 4~meV/at.~on total energy differences.
$NPT$ simulations change the volume of the supercell and therefore imply the presence of Pulay stresses if plane-wave basis sets are used~\cite{DaCosta1986}.
In our calculations, we estimate that the absolute value of the volume is systematically underestimated by roughly 0.5\% with respect to static calculations of the equilibrium volume with the Birch-Murnaghan equation of state~\cite{Murnaghan1944,Birch1947}.
The structural relaxations at 0~K have been performed on both the atomic and lattice degrees of freedom until all the forces were less than 0.01~eV/\AA~and all the components of the stress tensor were less than 100~MPa.
\newline
The simulations have been carried out in  $(4\times4\times4)$ supercells of the conventional orthorhombic cell of $\alpha''$ containing 256 atoms (see Fig.~\ref{SQS}).
The occupations of lattice sites have been determined according to special quasirandom structures (SQS) configurations~\cite{Zunger1990,Wei1990} generated with the Monte Carlo algorithm of a modified version~\cite{Pezold2010,Kossmann2015} of the ATAT package~\cite{VanDeWalle2002}.
In the minimization algorithm, geometrical correlations of pair, 3-body, 4-body, and 5-body figures have been considered up to the 9th, 5th, 4th, and 2nd neighbor shells, respectively.

\section{0 K Minimum Energy Path}

\begin{figure}[b]
\begin{centering} 
\includegraphics[scale=0.12]{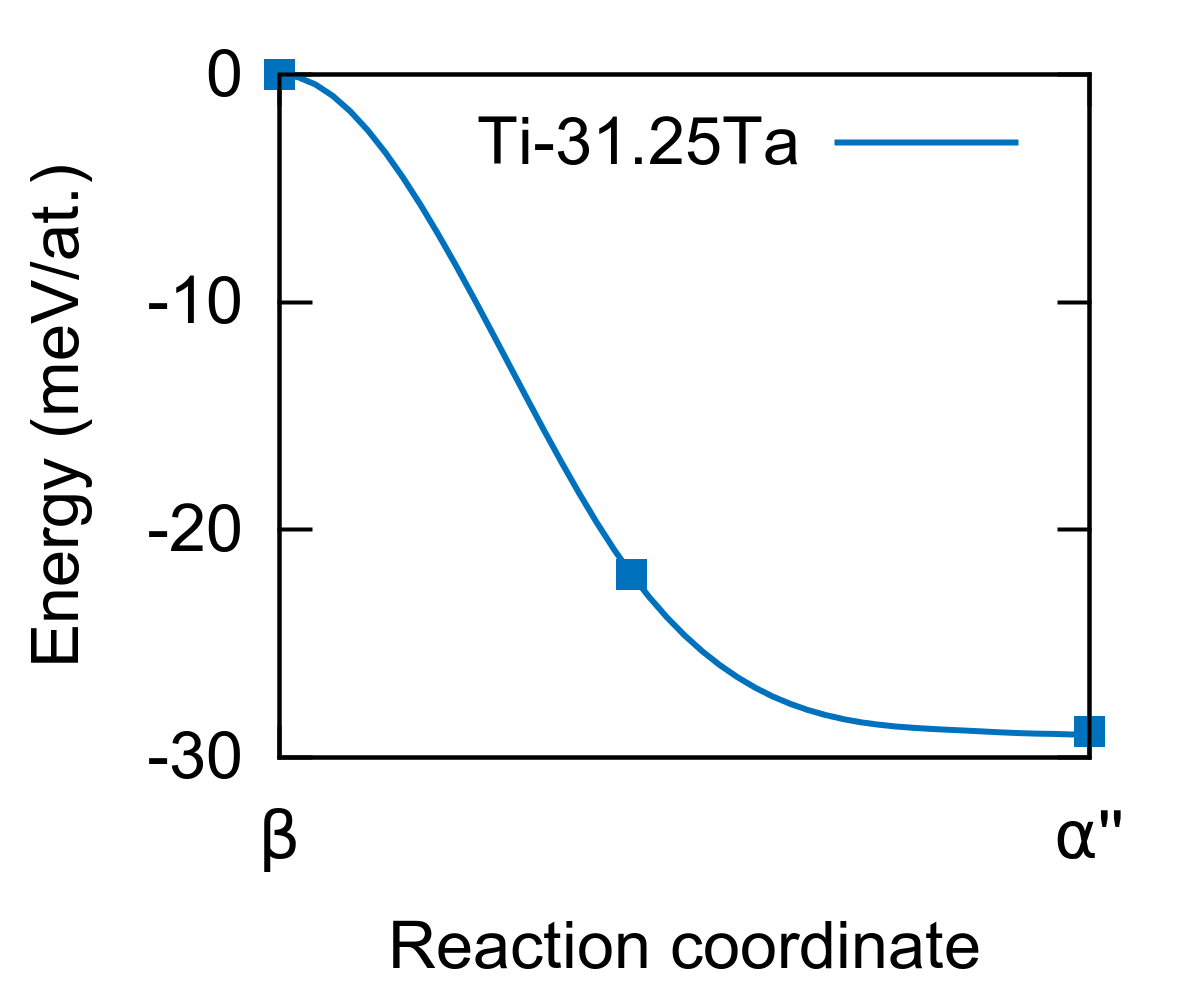}
\par\end{centering}
\caption{Minimum energy path at 0 K for the MT between austenite ($\beta$) and martensite ($\alpha''$) for Ti-31.25Ta.}
\label{neb}
\end{figure}

Fig.~\ref{neb} shows the minimum energy path for the MT in Ti-31.25Ta at 0 K obtained using the solid state nudged elastic band (SSNEB) method~\cite{Sheppard2012} as implemented in the VTST package~\cite{ssneb}.
The atomic positions in the austenitic phase have been determined using the average positions of an aiMD run at 600 K.
In agreement with previous calculations~\cite{Chakraborty2015}, the minimum energy path at 0 K does not display any barrier, meaning that static calculations are unable to capture even the first order nature of the MT.

\section{Transition Temperatures}
The transition temperatures  have been determined from the simulations by considering the temperature dependence of the spontaneous lattice strain of martensite (SLS) and the average atomic displacement ($\Delta y$) as a function of temperature.
For Ti-25Ta both order parameters drop to zero between 600 K and 650 K, while for Ti-31.25Ta at roughly 500 K.
By averaging the actual temperatures of the MD runs, the values for the transition temperature have been calculated to be $T_0=$ 627 K and 496 K for the two compositions, respectively.
\newline
The experimental transition temperatures have been evaluated as
\begin{equation}
T_0= \frac{M_\text{s}+A_\text{s}}{2}
\label{T0}
\end{equation}
\noindent
where $M_\text{s}$ and $A_\text{s}$ are  the martensitic and austenitic start temperatures, respectively.
We have taken the measured temperatures for Ti-Ta from Ref.~\cite{Ferrari2018} and linearly interpolated them to obtain $T_0$ values for the compositions Ti-25Ta and Ti-31.25Ta yielding 560~K and 420~K, respectively.

\section{Details on the Landau-Falk expansion}

\begin{table*}[t]
\caption{
\label{coeff_free_en}
\textbf{Columns 2-4}: input coefficients for the parametrization of the free energy.
\textbf{Columns 5-8}: coefficients of the Landau-Falk expansion.
\textbf{Columns 9-10}: conversion factors for the two order parameters.}
\begin{ruledtabular}
\begin{tabular}{cccc|cccc|cc}
 & $\Delta E^{(\beta - \alpha'')}$ (meV/at.) & $T_{0}$ (K) & $\eta_{0}$ & $a$ (meV/at.) & $b$ (meV/at.) & $c$ ($\mu$eV/at./K) & $T_{c}$ (K) & SLS$_{0}$ & $\Delta y_{0}$\tabularnewline
\hline
Ti-25Ta & 40 & 627 & 0.606 & 31.6 & 23.2 & 84.0 & 576 & 7.5 & 0.099\tabularnewline
Ti-31.25Ta & 29 & 496 & 0.536 & 20.3 & 11.7 & 79.3 & 475 & 6.0 & 0.093\tabularnewline
\end{tabular}
\end{ruledtabular}
\end{table*}

\begin{figure}[t]
\begin{centering} 
\includegraphics[scale=0.20]{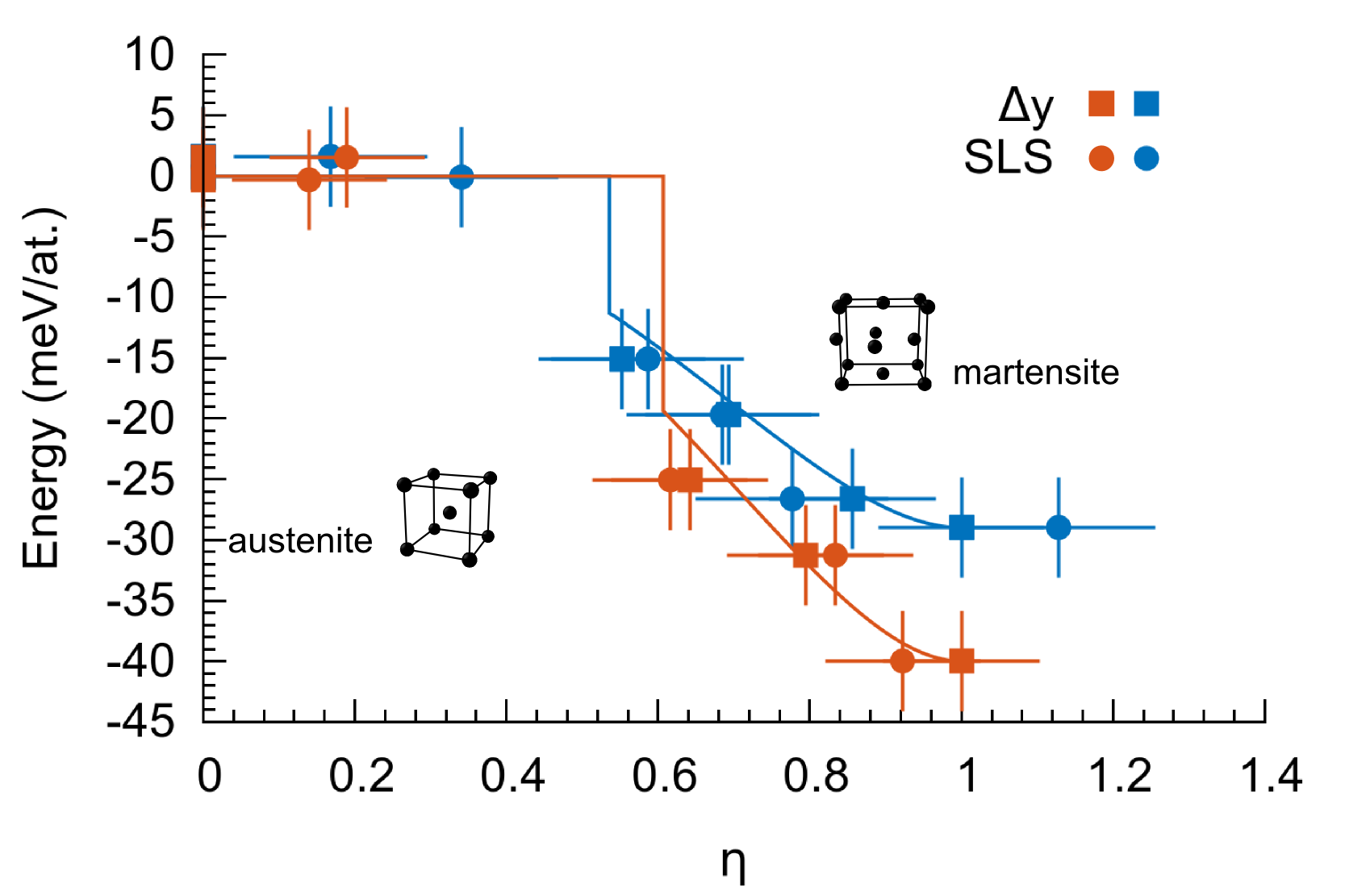}
\par\end{centering}
\caption{``Static'' contribution to the energy as a function of the order parameter for Ti-25Ta (red) and Ti-31.25Ta (blue).
The values of the order parameter are those which minimize the free energy at a given temperature and are obtained from the spontaneous lattice strain (circles), or from the average atomic displacements (squares).
Solid lines are predictions from the Landau-Falk expansion (no fit).
Since $\eta$ is normalized to 1, the numerical data for SLS and $\Delta y$ have been divided by the factors reported in the last two columns of Tab.~\ref{coeff_free_en}.}
\label{energy}
\end{figure}

\begin{figure}[b]
\begin{centering} 
\includegraphics[scale=0.16]{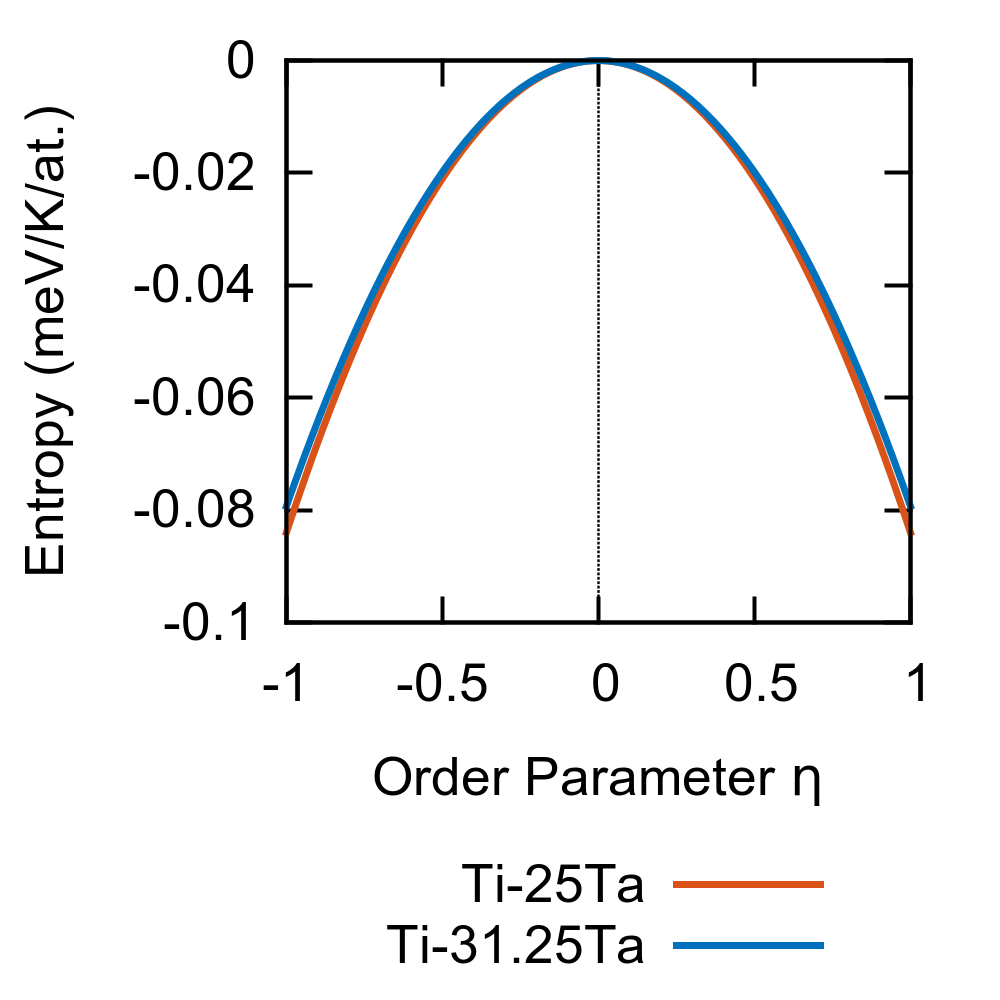}
\par\end{centering}
\caption{Entropy as a function of the order parameter for Ti-25Ta (red) and Ti-31.25Ta (blue) obtained from the Landau-Falk expansion.}
\label{entropy}
\end{figure}

\begin{figure}[b]
\begin{centering} 
\includegraphics[scale=0.20]{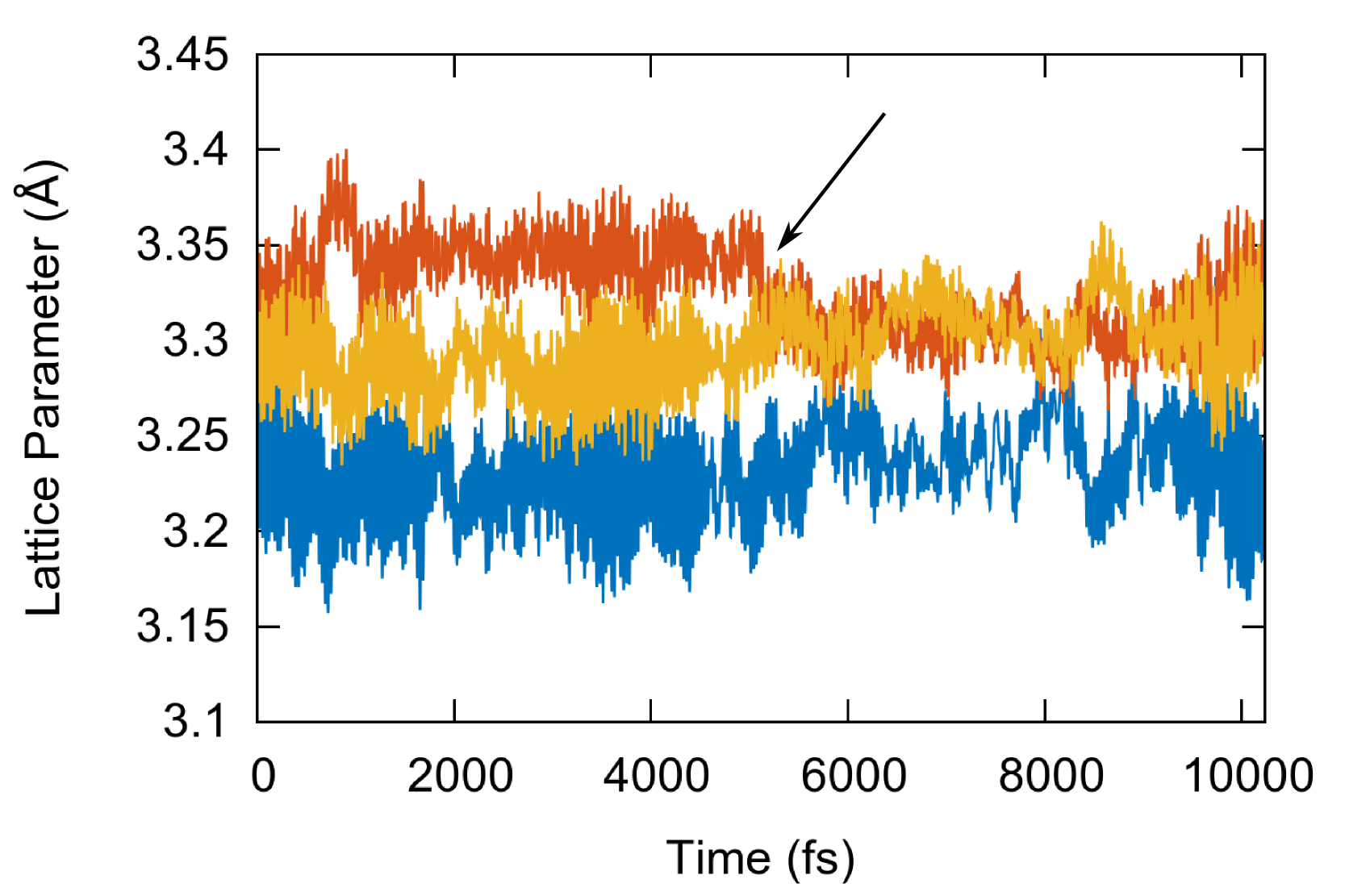}
\par\end{centering}
\caption{A 10 ps aiMD trajectory of Ti-31.25Ta at 500~K. The arrow points out the instant at which the MT takes place.}
\label{traj}
\end{figure}

In 1980, Falk \cite{Falk1980} proposed that the first order martensitic transformation (MT) in shape memory alloys can be described by the free energy

\begin{equation}
 F(\eta,T)=a \eta^6 - b \eta^4+ c (T-T_c) \eta^2+ F_0(T)
 \label{falk_free_en}
\end{equation}

\noindent
where $\eta$ is an order parameter, $a$, $b$, $c$, and $T_c$ are positive, material-dependent constants, and $F_0(T)$ decribes the temperature dependence of the absolute free energy of austenite. 
Without loss of generality, to treat the relative free energy difference between austenite and martensite, we have chosen $F_0(T)=0$.
\newline
To determine the values of $a$, $b$, $c$, and $T_c$ for Ti-25Ta and Ti-31.25Ta we have imposed the following conditions:
\begin{itemize}
 \item at 0 K, $F(\eta,T)$ has two minima at respectively $\eta=-1$ and $\eta=+1$;
 \item at 0 K, $F(0,0)-F(1,0)=\Delta E^{(\beta - \alpha'')}$, where $\Delta E^{(\beta - \alpha'')}$ is the 0~K energy difference between austenite and martensite;
 \item at the transition temperature $T_0$, $F(\eta,T_0)$ has two minima at respectively $\eta=-\eta_0$ and $\eta=+\eta_0$, where $\eta_0$ is the value of the order parameter at $T_0$ extracted from the $NPT$ simulations;
 \item at the transition temperature $T_0$, $F(0,T_0)-F(\eta_0,T_0)=0$.
\end{itemize} 

\noindent
To compute the total energy of austenite at 0 K, we have employed the average positions of the aiMD run at 700~K for Ti-25Ta and at 600~K for Ti-31.25Ta.
In fact, the chemical disorder in Ti-Ta implies that in the austenitic phase the average atomic positions do not correspond exactly to the perfect bcc positions.
\newline
The input parameters and the values of the coefficients of the free energy expansion are compiled in Tab.~\ref{coeff_free_en}.
\newline
The analytical values of the order parameter $\eta$ as a function of temperature can be derived from Eq.~\eqref{falk_free_en} by imposing

\begin{equation}
 \frac{\partial F(\eta,T)}{\partial\eta}=0.
 \label{ord_par_eq}
\end{equation}

\noindent
$\eta$ can be used to obtain the values of SLS and $\Delta y$ as a function of temperature, as done in Fig.~3 in the main text. 
Since $\eta$ is normalized to 1, the multiplicative factors listed in the last two colums of Tab.~\ref{coeff_free_en} have been used for the comparison of the analytical predictions of the Landau-Falk expansion to the simulation data for SLS and $\Delta y$.
\newline
To further test the validity of Eq.~\eqref{falk_free_en} for the free energy of our system, we have also extracted the values of the total energy as a function of the order parameter from our model as

\begin{equation}
 E(\eta,T)=\frac{\partial(\beta F(\eta,T))}{\partial\beta}
 \label{energy_eq}
\end{equation}

\noindent
where $\beta=\frac{1}{k_\text{B} T}$ and $k_\text{B}$ is the Boltzmann constant.
Considering that we have neglected the temperature dependence of the absolute free energy of austenite by setting $F_0=0$, Eq.~\eqref{energy_eq} gives the ``static'' contribution to the energy, i.e.~the energy that a system with a given value of the order parameter would have at 0 K.
We have hence performed additional calculations of the 0 K energy of our system for different values of the order parameter and compared the results to the analytical trends. 
Figure~\ref{energy} displays $E(\eta)$, where $\eta$ is either normalized SLS or $\Delta y$.
The analytical predictions are in excellent agreement with the numerical results for both compositions, and the data for SLS and $\Delta y$ agree with each other very well.
The discontinuous jump in the energy is the latent heat $T_0 \Delta S$ of the MT. 
A finite latent heat also confirms the first order nature of the phase transition.
\newline
The entropy difference between austenite and martensite can be obtained from Eq.~\eqref{falk_free_en} as
\begin{equation}
 S(\eta,T)=-\frac{\partial F(\eta,T)}{\partial T}.
 \label{ord_par_eq}
\end{equation}
If $F_0(T)=0$, the entropy does not depend on temperature $S(\eta,T)=S(\eta)$.
As can be seen in Fig.~\ref{entropy}, where $S$ is plotted as a function of $\eta$, the entropy of the austenitic phase is higher than that of the martensitic phase.
This favors the austenitic phase over the martensitic phase at high temperatures.
Furthermore, the actual value of the entropy depends very weakly on the composition.
This  peculiar characteristic of Ti-Ta-based alloys  was already assumed in previous works~\cite{Chakraborty2015,Ferrari2018} on these materials, where the compositional dependence of the phase stability, which in general depends on both energy and entropy, has been correlated only to 0 K energy differences, supposing that the entropy difference is constant as a function of the chemical concentration.
\newline
The analytical model provides also the range of temperatures in which martensite and austenite are both stable (metastability region of the MT)

\begin{equation}
 \Delta T=\frac{b^2}{3ac}
 \label{hysteresis}
\end{equation}

\noindent
and the height of the barrier at the transition temperature $T_0$

\begin{equation}
 \Delta E_{\text{barr}}=-\frac{k\cdot\left[6ac\cdot(T_{0}-T_{c})+b\cdot k\right]}{27a^{2}}
 \label{barrier}
\end{equation}

\noindent
where

\begin{equation}
 k=-b+\sqrt{b^{2}+3ac\cdot(T_{c}-T_{0})}
 \label{barrier2}
\end{equation}

\noindent
The error bars associated with the values of $T_0 \Delta S$, $\Delta T$ and $\Delta E_\text{barr}$ have been determined by a sensitivity analysis.
The factor that influences the most the latent heat $T_0 \Delta S$ is $\Delta E^{(\beta - \alpha'')}$.
Deviations of 5 meV/at.~on this quantity change the latent heat by roughly 2-3 meV/at., hence a value of 3 meV/at.~has been taken as the absolute error in this case. 
The reported values for $\Delta T$ and $\Delta E_\text{barr}$, instead, have been found to be almost insensitive to variations of 5 meV/at.~and 50 K in the parameters $\Delta E^{(\beta - \alpha'')}$ and $T_0$. 
Changes of the order of 5\% on $\eta_0$ though affected the final value of these quantities by roughly 30\%.
This has therefore been assumed as the relative error on $\Delta T$ and $\Delta E_\text{barr}$.

\section{A Martensitic Transformation during the MD run}

\begin{figure}
\begin{centering} 
\includegraphics[scale=0.16]{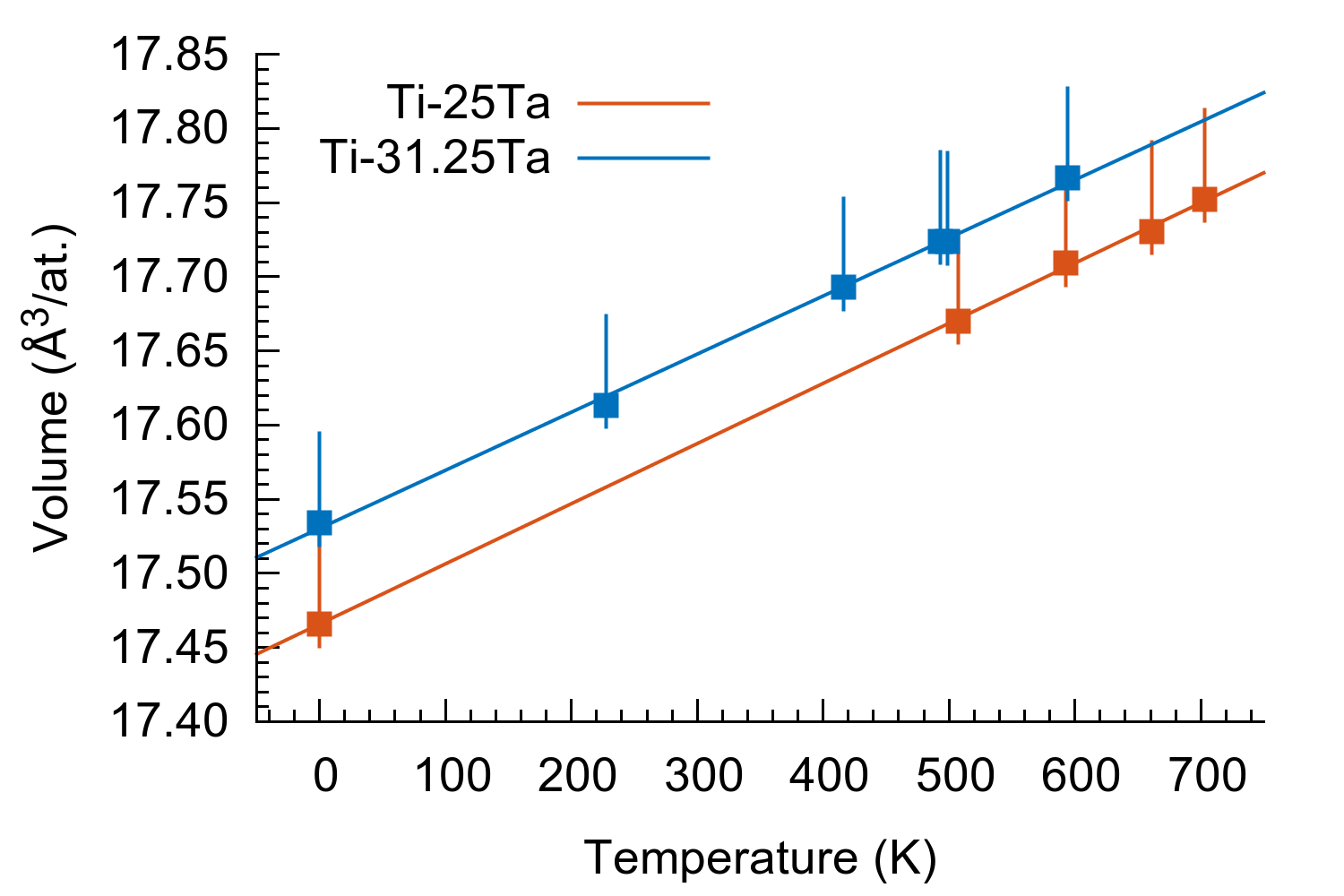}
\par\end{centering}
\caption{The atomic volume as a function of temperature for Ti-25Ta (red) and Ti-31.25Ta (blue). The asymmetric error bars take into account the Pulay stresses.}
\label{volume}
\end{figure}

For Ti-31.25Ta we detected a MT during a 10 ps aiMD run at 500 K (figure \ref{traj}): 
at $\simeq 5000$ fs the lattice parameters $b$ and $c$ become equal in magnitude and the system transforms from $\alpha''$ to $\beta$.
This supports the calculated value for the free energy barrier of Ti-31.25Ta at $T_0$ of 100~$\mu$eV/at.; 
indeed, a typical time scale for the detection of a MT can be estimated as 
\begin{equation}
 \tau \sim \frac{1}{\nu_0 e^{-\frac{\Delta E_\text{tot}}{k_\text{B}T_0}}}
\end{equation}
where $\Delta E_\text{tot}$ is the absolute barrier for the MT and $\nu_0$ is the attempt frequency.
For the concerted transformation of the entire system $\Delta E_\text{tot}$ scales with the system size, hence for 256 atoms $\Delta E_\text{tot}\simeq25.6$~meV. 
Assuming a value of approximately $10^{12}$~Hz for $\nu_0$, we obtain $\tau \sim 1000$~fs, in agreement with the time scale at which the MT takes place in the numerical simulations.
\bigskip

\section{Temperature Dependence of the Volume}
Simulations in the $NPT$ ensemble make it possible to compute the equilibrium volume of the system as a function of temperature.
Figure~\ref{volume} presents the average atomic volume extracted from the calculations as a function of temperature.
Despite the first order character of the MT, the atomic volume appears to be almost continuous before and after the MT.
This demonstrates that the $\alpha''$ and $\beta$ phases have approximately the same volume at the transition temperature.
The exceptionally small change of the volume at the transition temperature is one of the factors that favor high reversibility, in agreement with the small height of the barrier predicted by the free energy expansion.

\bibliography{./bib/biblio}